\documentclass{ifacconf}

\usepackage{graphicx}      % include this line if your document contains figures
\usepackage{natbib}        % required for bibliography
\usepackage{amsmath,amssymb}
\usepackage{lipsum}
\usepackage{pgfplots}
\usepackage{import}
\usepackage{caption}
\usepackage{subcaption}
\begin{document}
\begin{frontmatter}

\title{Disturbance feedback-based model predictive control in uncertain dynamic environments} 

\author[First]{Philipp Buschermöhle} 
\author[First]{Taouba Jouini}
\author[First]{Torsten Lilge} 
\author[First]{Matthias A. Müller}

\address[First]{Leibniz University Hannover, Institute of Automatic Control,
	 Hannover, Germany (E-mail: {buschermoehle,jouini,lilge,mueller}@ irt.uni-hannover.de).}

\begin{abstract}    
This paper presents a robust MPC scheme for linear systems subject to time-varying, uncertain constraints that arise from uncertain environments. The predicted input sequence is parameterized over future environment states to guarantee constraint satisfaction despite an imprecise environment prediction and unknown evolution of the future constraints. We provide theoretical guarantees for recursive feasibility and asymptotic convergence. 
Finally, a brief simulation example showcases our results.

\end{abstract}

\begin{keyword}
Model Predictive Control, Time-Varying Constraints, Disturbance Feedback
\end{keyword}

\end{frontmatter}
%===============================================================================

\section{Introduction}
Modern control applications often have high safety requirements. Conventionally, this revolves around keeping a controlled system within a safe operating region. During control design, such limitations are usually incorporated as state and input constraints. However, in recent years, the environments that control applications operate in become more and more complex raising new types of safety considerations. Especially when operating close to humans, the emphasis on safety gets even more important. Thus, applications such as autonomous driving in mixed traffic situations \citep{Batkovic23b} or human-machine cooperation in robotics \citep{Robla-Gomez17} demand special safety considerations. In those cases, the constraints on the system that ensure safety become time-varying and not exactly predictable.

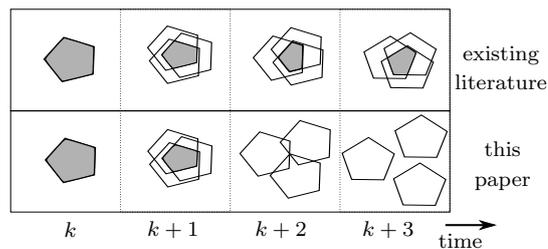
\begin{figure}
	\begin{center}
		{\footnotesize
		\def\svgwidth{0.8\linewidth}
		%% Creator: Inkscape 1.1.1 (3bf5ae0d25, 2021-09-20), www.inkscape.org
%% PDF/EPS/PS + LaTeX output extension by Johan Engelen, 2010
%% Accompanies image file 'Constraint_sets.pdf' (pdf, eps, ps)
%%
%% To include the image in your LaTeX document, write
%%   \input{<filename>.pdf_tex}
%%  instead of
%%   \includegraphics{<filename>.pdf}
%% To scale the image, write
%%   \def\svgwidth{<desired width>}
%%   \input{<filename>.pdf_tex}
%%  instead of
%%   \includegraphics[width=<desired width>]{<filename>.pdf}
%%
%% Images with a different path to the parent latex file can
%% be accessed with the `import' package (which may need to be
%% installed) using
%%   \usepackage{import}
%% in the preamble, and then including the image with
%%   \import{<path to file>}{<filename>.pdf_tex}
%% Alternatively, one can specify
%%   \graphicspath{{<path to file>/}}
%% 
%% For more information, please see info/svg-inkscape on CTAN:
%%   http://tug.ctan.org/tex-archive/info/svg-inkscape
%%
\begingroup%
  \makeatletter%
  \providecommand\color[2][]{%
    \errmessage{(Inkscape) Color is used for the text in Inkscape, but the package 'color.sty' is not loaded}%
    \renewcommand\color[2][]{}%
  }%
  \providecommand\transparent[1]{%
    \errmessage{(Inkscape) Transparency is used (non-zero) for the text in Inkscape, but the package 'transparent.sty' is not loaded}%
    \renewcommand\transparent[1]{}%
  }%
  \providecommand\rotatebox[2]{#2}%
  \newcommand*\fsize{\dimexpr\f@size pt\relax}%
  \newcommand*\lineheight[1]{\fontsize{\fsize}{#1\fsize}\selectfont}%
  \ifx\svgwidth\undefined%
    \setlength{\unitlength}{787.3620221bp}%
    \ifx\svgscale\undefined%
      \relax%
    \else%
      \setlength{\unitlength}{\unitlength * \real{\svgscale}}%
    \fi%
  \else%
    \setlength{\unitlength}{\svgwidth}%
  \fi%
  \global\let\svgwidth\undefined%
  \global\let\svgscale\undefined%
  \makeatother%
  \begin{picture}(1,0.44727948)%
    \lineheight{1}%
    \setlength\tabcolsep{0pt}%
    \put(0,0){\includegraphics[width=\unitlength,page=1]{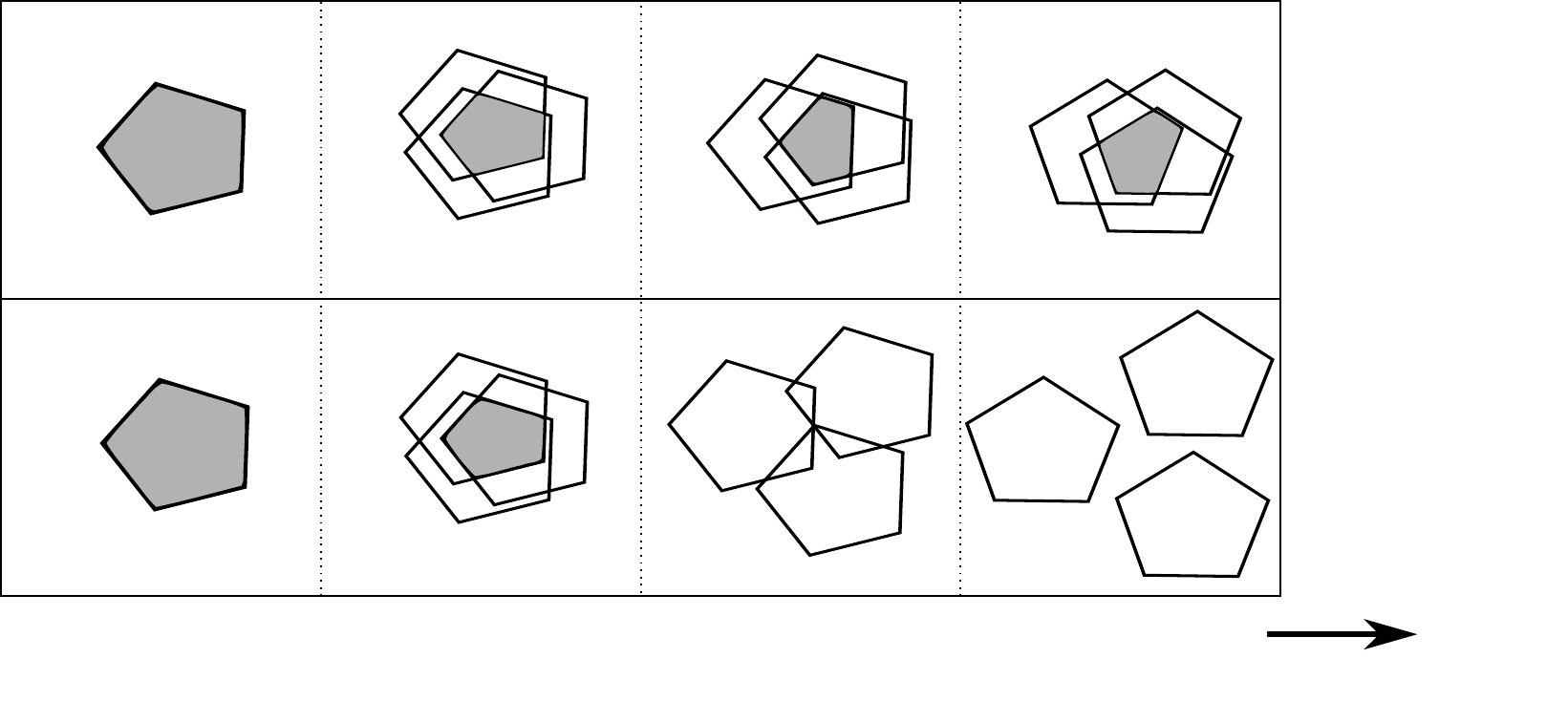}}%
    \put(0.83648794,0.00053953){\makebox(0,0)[t]{\lineheight{1.25}\smash{\begin{tabular}[t]{c}time\end{tabular}}}}%
    \put(0.91203795,0.35086466){\makebox(0,0)[t]{\lineheight{1.25}\smash{\begin{tabular}[t]{c}existing \\literature\end{tabular}}}}%
    \put(0.91481049,0.17091459){\makebox(0,0)[t]{\lineheight{1.25}\smash{\begin{tabular}[t]{c}this\\paper\end{tabular}}}}%
    \put(0.11200491,0.02129251){\makebox(0,0)[t]{\lineheight{1.25}\smash{\begin{tabular}[t]{c}$k$\end{tabular}}}}%
    \put(0.30270528,0.02152884){\makebox(0,0)[t]{\lineheight{1.25}\smash{\begin{tabular}[t]{c}$k+1$\end{tabular}}}}%
    \put(0.50284561,0.02198759){\makebox(0,0)[t]{\lineheight{1.25}\smash{\begin{tabular}[t]{c}$k+2$\end{tabular}}}}%
    \put(0.70231308,0.02221705){\makebox(0,0)[t]{\lineheight{1.25}\smash{\begin{tabular}[t]{c}$k+3$\end{tabular}}}}%
  \end{picture}%
\endgroup%
}
		\caption{{\small An example of the uncertain evolution of constraint sets predicted at time $k$ in existing literature (e.g. \citep{Liu19}) and this paper. The intersection of the predicted constraint sets is depicted in gray.}}
		\label{fig:constr}
		\end{center}
	\end{figure}

While there exists a large variety of algorithms to handle constraints, such as reference governors \citep{Kolmanovsky14} or control barrier functions \citep{Ames19} just to name two, model predictive control is of special interest as it is one of the most widely used algorithms that allows intuitive constraint integration \citep{Mayne14}. The case of time-invariant constraints has been thoroughly investigated with conditions for recursive feasibility and stability. Similar results exist for the case of time-varying, but known constraints \citep{Rawlings17,Mayne09,Manrique14}. For a priori unknown, time-varying constraints, there are fewer results. In \citep{Liu19}, an MPC scheme for time-varying, uncertain polytopic constraints is presented by including tightened constraints that guarantee satisfaction of any possible future constraint. Recursive feasibility is achieved by a fixed terminal region that is always feasible after a certain time. In \citep{Batkovic23}, a trajectory tracking MPC scheme for uncertain environments has been developed. The authors include the uncertain environment via time-varying uncertain, predicted inequalities. The set that is defined by those inequalities is assumed to increase over time. Recursive feasibility results here from the combination of this assumption on the constraints and a terminal set that is based on a safe set. \cite{Soloperto19} combine an uncertain, set-based prediction of an obstacle with a robust MPC scheme for collision avoidance. Obstacle avoidance is guaranteed by avoiding the predicted set of future obstacle states in the constraints of the MPC. These results share that the evolution of the uncertain constraints is limited in such a way that, at any prediction step, the intersection of the possible constraint sets is non-empty. Thus, there is always one single feasible trajectory that satisfies the constraints. In our work, we consider more general cases where the intersection of the constraints resulting from different environment states may be empty, compare Fig.~\ref{fig:constr}.  

While robust MPC (RMPC) approaches for uncertain prediction models or additive process disturbances have been thoroughly investigated and lead to successful approaches such as tube MPC 
\citep{Langson04} or disturbance feedback MPC (DFMPC) \citep{Goulart06}, the case of uncertain constraints has not been as thoroughly investigated. \cite{Nair22} formulate robust and stochastic MPCs for avoidance of an uncertain dynamic obstacle by introducing a feedback term from the uncertain obstacle to the controlled system. However, the authors provide no results on recursive feasibility and stability. ßcite{Bonzanini24} address the problem of uncertain constraints in a stochastic setting and provide probabilistic guarantees.

In our work, we apply a similar rationale as \citep{Nair22} and employ a feedback from the environment to the system with an input parameterization that is similar to the parameterization of DFMPC \citep{Goulart06} while considering uncertain, polytopic constraints. Optimizing over the feedback from future environment states allows us to consider a more general setting that can handle larger uncertainty of the environment and, thus, larger uncertainty of the constraints including non-intersecting constraints as depicted in Fig.~\ref{fig:constr}. Moreover, we present a prediction of the uncertain environment that can be included in an MPC algorithm. Our main contribution consists in providing guarantees for recursive feasibility and asymptotic convergence of the state under sufficient conditions. Finally, we demonstrate the proposed scheme in a simulation example.
\subsection{Notation}
For a time-dependent variable, 
$x(k)$
denotes the actual values of the respective variables at time $k$ and  
$x_{i|k}$ 
denotes the $i$-step ahead prediction of this variable made at time $k$. 
We define the set of integers between $a,b\in\mathbb{Z}$ with $a\leq b$ by $\mathbb{Z}_{[a,b]}:=\{z\in\mathbb{Z}\,|\,a\leq z\leq b\}$ and the set of non-negative real numbers by $\mathbb{R}_{\geq 0}$. For a vector $x$ or a matrix $A$, $x^\top$ and $A^{\top}$ are the transposes and $[x]_j$ and $[A]_j$ denote the $j$-th element of a vector and the $j$-th row of a matrix, respectively. We use $||x||$ and $||A||$ for an arbitrary $p$-Norm and the respective induced matrix norm. A continuous function $\alpha:\mathbb{R}_{\geq 0}\rightarrow\mathbb{R}_{\geq 0}$ is a class $\mathcal{K}$-function if $\alpha(0)=0$ and $y>x\implies \alpha(y)>\alpha(x)$. If, additionally, $\alpha $ is radially unbounded, $\alpha$ is a $\mathcal{K}_{\infty}$-function. 
\section{Preliminaries and setup}
% ---------------------------------------------------------------------
\subsection{Problem formulation}
Consider the following discrete-time LTI system
\begin{eqnarray}
	x^+= Ax+Bu
	\label{equ:DynEq}
\end{eqnarray}
with state and input $x\in \mathbb{R}^{n_x}$, $u\in\mathbb{R}^{m_x}$ and system matrices $A\in \mathbb{R}^{n_x\times n_x}$, and $B\in \mathbb{R}^{n_x\times m_x}$. The pair $(A,B)$ is stabilizable. The goal is to drive the system state to the origin while satisfying time-varying, a priori unknown polytopic state constraints $\mathcal{X}(k)$ and time-invariant, polytopic input constraints $\mathcal{U}$. The state constraints depend on an uncertain environment state $o\in\mathbb{R}^{n_o}$ that evolves according to the known linear dynamics
\begin{eqnarray}
	o^+= To+Sv
	\label{equ:DynEqEnv}
\end{eqnarray}
with environment input $v\in\mathbb{R}^{m_o}$, and system matrices $T\in \mathbb{R}^{n_o\times n_o}$, and $S\in \mathbb{R}^{n_o\times m_o}$. Let the state constraint set be given by
\begin{equation}
	\mathcal{X}(k) := \left\{ x\in \mathbb{R}^{n_x}\,|\, Fx+Go(k)\leq g\right\},
\end{equation}
with constant $F\in\mathbb{R}^{\nu \times n_x}$, $G\in\mathbb{R}^{\nu \times n_o}$, and $g\in\mathbb{R}^{\nu}$. There is further a compact set $\mathcal{X}$ such that $\mathcal{X}(k)\subseteq \mathcal{X}$ for all times $k$. 
Additionally, we assume knowledge on the future environment that is specified by two ingredients. (I) The environment does not change arbitrarily fast in the future and (II) the environment can be roughly bounded%, and (III) the environment admits a region around the origin after some finite time
. The latter allows for unstable or marginally stable environments. This is especially interesting if there is no exact model of the environment, but only limits on the environment state and its rate of change are known, which makes an integrator a suitable environment model. We describe these two parts in the following and combine them to a prediction of environment trajectories.

(I) At time $k$, polytopic prediction sets of the future environment input $\mathcal{V}_{i|k} $ are known for all $i=0...N-1$ and given by
\begin{equation}
	 \mathcal{V}_{i|k}:=\left\{v\,|\, E_{i|k}v\leq e_{i|k}\right\}.\label{equ:v_tube}
\end{equation} 
Further, there is a compact set $\mathcal{V}$ such that $\mathcal{V}_{i|k}\subseteq \mathcal{V}$ and $||v||\leq \bar{v}$ for all $v\in\mathcal{V}$.

(II) At time $k$, polytopic prediction sets of the environment state are known for all $0\leq i \leq N$ and given by $\mathcal{O}_{i|k}$ such that
\begin{equation}
	\mathcal{O}_{i|k}:=\left\{o\,|\, H_{i|k}o\leq h_{i|k}\right\}.\label{equ:outer_tube}
	\end{equation} 
We assume that the actual future environment state $o(k+i)$ and input $v(k+i)$ are contained by $\mathcal{O}_{i|k}$ and $\mathcal{V}_{i|k}$,
	\begin{align}
		&v(k+i)\in\mathcal{V}_{i|k}\\
		&o(k+i)\in\mathcal{O}_{i|k}.
	\end{align}

\begin{rem}\label{rem:ex}
	To illustrate the description of the environment, consider the example 
	of a human-controlled vehicle with states $o$. The inputs $v$ are the jerk and steering angle. We can determine a maximally physically admissible set $\mathcal{V}$. Depending on the traffic situation, tighter sets can be determined for the prediction. For example, if the vehicle $o$ stands in front of a red light, in the next few steps, the set $\mathcal{V}_{i|k}$ only contains a small set of jerks around zeros, whereas for further prediction steps only nonnegative jerks are predicted as the traffic light may turn green. The sets $\mathcal{O}_{i|k}$ can, e.g., consist of limits on the velocity derived from speed limits or on the position based on the border of the road. For a detailed description of a set-based prediction of traffic participants, we refer the reader to \cite{Althoff16}.
\end{rem}

We generate the prediction of the environment by combining the sets $\mathcal{V}_{i|k}$ and $\mathcal{O}_{i|k}$. We construct a set of admissible trajectories for $o$, given by $\boldsymbol{\mathcal{O}_{k}}$, by using the environment dynamic \eqref{equ:DynEqEnv} as follows
\begin{equation}
	\boldsymbol{\mathcal{O}_{k}}=\left\{ \left. \begin{pmatrix}
		o_{0|k}\\\vdots\\o_{N|k}
	\end{pmatrix}\right|\begin{array}{l}
	o_{i|k} \in\mathcal{O}_{i|k},\\
	o_{i|k}-To_{i-1|k}\in S\mathcal{V}_{i-1|k}\\
	\forall i\in\mathbb{Z}_{[1,N]},\quad o_{0|k} = o(k)\\
\end{array}\right\}.\label{equ:O_pred}
\end{equation}
The set $\boldsymbol{\mathcal{O}_{k}}$ is polyhedral because it is constructed by concatenating and intersecting polyhedral sets. From \eqref{equ:O_pred}, it can be seen that the dynamics \eqref{equ:DynEqEnv} and the sets $\mathcal{V}_{i|k}$ describe the expansion of the environment states while the sets $\mathcal{O}_{i|k}$ are additional bounds that limit the uncertainty expansion.

\subsection{Optimal control problem}
The control objective is to steer system \eqref{equ:DynEq} to the origin while satisfying the constraints $x(k)\in\mathcal{X}(k)$ that depend on the unknown environment state. Instead of predicting one input sequence as in \cite{Liu19}, we use a parameterized control law to handle larger uncertainties of the constraint sets as depicted in Fig.~\ref{fig:constr}. Hence, we modify the approach from \cite{Goulart06} and use a control policy that is the sum of a nominal input $c_{i|k}$ and an environment feedback term $\sum_{l=0}^{i}K_{(i,l)|k}o_{i|k}$, where $K_{(i,j)|k}\in\mathbb{R}^{m_x\times n_o}$ are environment feedback gain matrices,
\begin{equation}
	u_{i|k} = c_{i|k}+\sum_{l=0}^{i}K_{(i,l)|k}o_{i|k}.
	\label{equ:inputParam}
\end{equation}
The proposed input parameterization yields input and state trajectories that depend on the future environment.

We define $\boldsymbol{c_k}:=(c_{0|k}^T,c_{1|k}^T,\dots,c_{N-1|k}^T)^T$ as the concatenated vector of nominal inputs and $\boldsymbol{K_k}\in\mathbb{R}^{Nm_x \times (N+1) n_o}$ as the feedback gain block matrix
\begin{equation}
	\hspace{-1ex}\boldsymbol{K_k}=\begin{pmatrix}
		K_{(0,0)|k}&0&\dots &0&0\\
		K_{(1,0)|k}&K_{(1,1)|k}&\dots &0&0\\
		%K_{(2,0)|k}&K_{(2,1)|k}&K_{(2,2)|k}&\dots&0\\
		%K_{(3,0)|k}&K_{3,1}&K_{3,2}&\dots &0\\
		\vdots &\vdots&\ddots &\vdots &\vdots\\
		K_{(N-1,0)|k}&K_{(N-1,1)|k}&\dots & K_{(N-1,N-1)|k}&0
	\end{pmatrix}\hspace{-0.75ex}.\hspace{-1ex}
\end{equation} 
In summary, we can write the predicted input sequence in a compact way as
\begin{equation}
	\boldsymbol{u_k} = \boldsymbol{c_k}+\boldsymbol{K_k}\boldsymbol{o_k}\label{equ:uVec}
\end{equation}
with the boldletter variables being the concatenation of all prediction steps of the respective variable, $\boldsymbol{u_k}:=(u_{0|k}^T,u_{1|k}^T,\dots,u_{N-1|k}^T)^T$ and $\boldsymbol{o_k}:=(o_{0|k}^T,o_{1|k}^T,\dots,o_{N|k}^T)^T$. The arising state sequence $\boldsymbol{x_k}:=(x_{0|k}^T,x_{1|k}^T,\dots,x_{N|k}^T)^T$ with an initial state $x$ can be determined to
\begin{equation}
	\boldsymbol{x_k}=\boldsymbol{A}x+\boldsymbol{B}\boldsymbol{u_k}=\boldsymbol{A}x+\boldsymbol{B}\boldsymbol{c_k}+\boldsymbol{B}\boldsymbol{K_k}\boldsymbol{o_k}\label{equ:xVec}
\end{equation}
with the matrices
\begin{equation}
	\boldsymbol{A}=\begin{pmatrix}
		I\\A\\A^{2}\\\vdots\\A^{N}
	\end{pmatrix}\,\textup{and } \boldsymbol{B}=\begin{pmatrix}
		0&0&\dots&0\\B&0&\dots&0\\AB&B&\dots&0\\ \vdots&\vdots &\ddots&\vdots\\A^{N-1}B&A^{N-2}B&\dots &B
	\end{pmatrix}.
\end{equation}

Following \cite{Goulart06}, we can formulate the set of feasible control policies for some initial state $x$ as
\begin{equation}
	\hspace{-1ex}\Pi_k^{c,K}(x):=\hspace{-0.5ex}\left\{\hspace{-0.5ex}(\boldsymbol{c_k},\boldsymbol{K_k})\begin{array}{l|ll}
		 \hspace{-1ex}&&x_{i+1|k}=Ax_{i|k}+Bu_{i|k}\\
		 \hspace{-1ex}&&u_{i|k}=c_{i|k} + \sum\limits_{l=0}^{i}K_{(i,l)|k}o_{j|k}\\
		 \hspace{-1ex}&&Fx_{i|k}+Go_{i|k}\leq g,\, u_{i|k}\in\mathcal{U}\\
		 \hspace{-1ex}&&\forall i=0,\dots,N-1\\
		 \hspace{-1ex}&&\forall \boldsymbol{o_k}\in \boldsymbol{\mathcal{O}_{k}}\\
		 \hspace{-1ex}&&x_{N|k}\in \mathcal{X}_f,\, x_{0|k}=x
	\end{array} \hspace{-0.5ex}\right\}.\hspace{-1ex}
\label{equ:paramSet}
\end{equation}

Here, $\mathcal{X}_f$ denotes a terminal region that we will specify later. Any policy in $\Pi_k^{c,K}$ yields a feasible state trajectory for any environment trajectory in $\boldsymbol{\mathcal{O}_{k}}$. Note that $\Pi_k^{c,K}(x)$ is time-varying due to the dependency of $\boldsymbol{\mathcal{O}_{k}}$ on the environment state $o(k)$ and the predictions $\mathcal{V}_{i|k}$ and $\mathcal{O}_{i|k}$ at time $k$. 

Based on these preliminaries, we can formulate an optimal control problem (OCP) that minimizes an appropriate cost function $J(\boldsymbol{x},\boldsymbol{u})$ while satisfying the constraints,
\begin{subequations}
	\begin{align}
		\min\limits_{\boldsymbol{c_k},\boldsymbol{K_k}}&&& J(\boldsymbol{\hat{x}_k},\boldsymbol{\hat{u}_k})\label{OCP:cost}\\
		\textup{s.t.} &&& \boldsymbol{\hat{x}_k}=\boldsymbol{A}x(k)+\boldsymbol{B}\boldsymbol{\hat{u}_k},\label{OCP:dyn}\\
		&&&\boldsymbol{\hat{u}_k}=\boldsymbol{c_k}+\boldsymbol{K}\boldsymbol{\hat{o}_k}\label{OCP:input}\\
		&&&(\boldsymbol{c_k},\boldsymbol{K_k})\in\Pi_k^{c,K}(x(k))\label{OCP:constr}.
	\end{align}
	\label{OCP}
\end{subequations}
In problem \eqref{OCP}, we choose one nominal admissible environment sequence $\boldsymbol{\hat{o}_k}\in\boldsymbol{\mathcal{O}_k}$ to construct the cost function in \eqref{OCP:cost}-\eqref{OCP:input} while ensuring feasibility for all $\boldsymbol{o_k}\in\boldsymbol{\mathcal{O}_{k}}$ via \eqref{OCP:constr}. The nominal input and state trajectories resulting from $\boldsymbol{\hat{o}_k}$ are given by $\boldsymbol{\hat{u}_k}$ and $\boldsymbol{\hat{x}_k}$. The cost function is given by the sum of a stage cost function $\ell:\mathbb{R}^{n}\times \mathbb{R}^{m}\rightarrow \mathbb{R}_{+}$ and a terminal cost function $V_f:\mathbb{R}^{n}\rightarrow \mathbb{R}_{+}$, i.e.
\begin{equation}
	J(\boldsymbol{\hat{x}_k},\boldsymbol{\hat{u}_k})=V_f(\hat{x}_{N|k})+\sum\limits_{i=0}^{N-1}\ell (\hat{x}_{i|k},\hat{u}_{i|k}).\label{equ:nomCost}
\end{equation}
We denote the solution of \eqref{OCP} at time $k$ by $\boldsymbol{c_k^{\ast}}$ and $\boldsymbol{K_k^{\ast}}$ and their elements by $c_{i|k}^{\ast}$ and $K_{(i,j)|k}^{\ast}$, respectively. We compute the solution of \eqref{OCP} at each time $k$ and apply only the first input $u_{0|k}^{*}$ to the system, such that
\begin{equation}
	u(k) = c_{0|k}^{\ast}+K_{(0,0)|k}^{\ast}o(k).\label{equ:inputAtk}
\end{equation}
If the solution is not unique, an arbitrary minimizer is chosen.
\begin{rem}
	\emph{(Disturbance-based parameterization)} Alternatively to \eqref{equ:inputParam}, we can paramterize the input using the inputs to the environment state,
	\begin{equation}
		u_{i|k} = \mu_{i|k}+\sum_{l=0}^{i-1}M_{(i,l)|k}v_{i|k},
		\label{equ:inputParam2}
	\end{equation} 
	where the nominal part of the input is denoted by $\mu_{i|k}$ and the feedback matrices by $M_{(i,l)|k}$. For this parameterization, previous values of $v_{k+i}$ need to be measured or reconstructed from the states. The OCP that arises from \eqref{equ:inputParam2} is constructed using the set of admissible policies $\Pi_k^{\mu,M}(x)$ based on a set of admissible environment inputs $\boldsymbol{\mathcal{V}_{k}}$. The details of the construction of $\boldsymbol{\mathcal{V}_{k}}$ and $\Pi_k^{\mu,M}(x)$ are omitted at this point as they are similar to \eqref{equ:O_pred} and \eqref{equ:paramSet}. It is easy to show that the optimization problem for \eqref{equ:inputParam2} is equivalent to the OCP \eqref{OCP} using similar arguments as in \citep{Goulart06}. However, in contrast to \citep{Goulart06}, both parameterizations are convex. Thus, there is no significant advantage using one over the other apart from a different number of decision variables depending on $n_o$ and $m_o$.
\end{rem}
\section{Main Results}
Before we formulate our results for the constraint satisfaction and stability of the proposed MPC scheme, we first derive an auxiliary result that is later used to show recursive feasibility of \eqref{OCP}. We require the following assumptions on the prediction of the environment.

\begin{assum}\label{assum:containedPred}
	The sets for the prediction of the environment decrease as $k$ increases, i.e.
	\begin{equation}
		\begin{aligned}
			&\mathcal{V}_{i-1|k+1}\subseteq\mathcal{V}_{i|k} \,&&\forall\, i\in\mathbb{Z}_{[1,N-1]}\\
			&\mathcal{O}_{i-1|k+1}\subseteq\mathcal{O}_{i|k}\,&&\forall\, i\in\mathbb{Z}_{[1,N]}\\
		\end{aligned}
	\end{equation}
	\end{assum}

\begin{assum}\label{assum:termRegion} There exists a compact set $\mathcal{X}_f$ that contains the origin in its interior and a control law $u=\kappa x$ such that $\kappa \mathcal{X}_f\subseteq \mathcal{U}$ and
	\begin{equation}
		x\in\mathcal{X}_f\implies (A+B\kappa )x\in\mathcal{X}_f.
	\end{equation}
\end{assum}
\begin{assum}\label{assum:termSetCont} There is a set $\mathbb{O}_f$ such that
	\begin{equation}
		\begin{aligned}
			\mathcal{O}_{N|k}\subseteq \mathbb{O}_f\subseteq \{o\,|\, Fx+Go\leq g\quad\forall\; x\in\mathcal{X}_f\}
		\end{aligned}
	\end{equation}
for all $k\geq 0$.
\end{assum}
While Assumption \ref{assum:termRegion} is a standard MPC assumption \citep{Rawlings17}, Assumptions \ref{assum:containedPred} and \ref{assum:termSetCont} are specific to our setting of uncertain environments. 
For the prediction of the environment state at time $k+i$, Assumption \ref{assum:containedPred} means that the uncertainty of the $(i-1)$-step ahead prediction of the environment at time $k+1$ is not larger than that of the $i$-step ahead prediction at time $k$.
Simply said, the prediction of the environment gets better (or stays equal) over time. This is reasonable in many applications. For example, constant sets $\mathcal{V}_{i|k}$ and $\mathcal{O}_{i|k}$ such as the traffic laws in Rem.~\ref{rem:ex} satisfy this assumption. Conceptually, this is similar to Assumption 3 in \cite{Soloperto19} and Assumption 4 in \cite{Batkovic23} that the predicted constraint set increases in size.
Assumption \ref{assum:termSetCont} is required since our control objective is setpoint stabilization. In particular, we require that for all times the terminal region is feasible at the end of the prediction horizon, i.e. that the environment is such that the resulting state constraints include a region around the origin. An analogous assumption is also made in \citep{Liu19}.

The following proposition shows that the set of possible environment trajectories predicted at time $k+1$ is contained in the set of possible environment trajectories predicted at time $k$, which intuitively means that the prediction improves over time.
\begin{prop} \label{prop:ContPred}
	\emph{(Shrinking environment prediction)} Let Assumptions \ref{assum:containedPred} and \ref{assum:termSetCont} hold, then 
\begin{equation}
	\{o(k)\}\times \boldsymbol{\mathcal{O}_{k+1}}\subseteq \boldsymbol{\mathcal{O}_{k}}\times \mathbb{O}_f.
\end{equation}
\end{prop}
\begin{pf} 
	Consider any sequence 
	\begin{equation*}
	\begin{pmatrix}
		o^\top(k)&o_{0|k+1}^\top&\dots &o_{N|k+1}^\top
	\end{pmatrix}^\top\in \{o(k)\}\times \boldsymbol{\mathcal{O}_{k+1}}.
\end{equation*} From the construction of $\boldsymbol{\mathcal{O}_{k+1}}$ and Assumptions \ref{assum:containedPred} and \ref{assum:termSetCont}, it directly follows that 
\begin{equation*}
	o_{N|k+1}\in\mathcal{O}_{N|k+1}\subseteq\mathbb{O}_f,
\end{equation*}
\begin{equation}
	o_{i-1|k+1}\in\mathcal{O}_{i-1|k+1}\subseteq \mathcal{O}_{i|k}\quad \forall \, i\in\mathbb{Z}_{[2,N]},
	\label{equ:pf_env}
\end{equation}
\begin{equation}
	o_{i|k+1}-To_{i-1|k+1}\in S\mathcal{V}_{i-1|k+1}\subseteq S\mathcal{V}_{i|k}\quad\forall \, i\in\mathbb{Z}_{[1,N-1]},
	\label{equ:pf_dyn}
\end{equation} 
\begin{equation*}
\mathrm{and}\;o_{0|k+1}=o(k+1).
\end{equation*}
Per assumption on the environment $o(k+1)\in\mathcal{O}_{1|k}$ follows. With \eqref{equ:pf_env}, $o_{i-1|k+1}\in\mathcal{O}_{i|k}$ holds for all $i\in\mathbb{Z}_{[1,N]}$. Because $o_{0|k+1}=o(k+1)=To(k)+Sv(k)=To(k)+Sv(k)$ and $v(k)\in\mathcal{V}_{0|k}$, together with \eqref{equ:pf_dyn}, this
now implies $\begin{pmatrix}
	o^\top(k)&o_{0|k+1}^\top&\dots &o_{N-1|k+1}^\top
\end{pmatrix}^\top\in\boldsymbol{\mathcal{O}_{k}}$ which concludes the proof. $\hfill\square$
\end{pf}
\begin{rem}
	Instead of using a fixed terminal region $\mathcal{X}_f$, we could alternatively utilize safe sets as described in \citep{Batkovic23} to achieve recursive feasibility and safe operation. In applications that do not require setpoint stabilization, this may yield a more reasonable assumption than Assumption \ref{assum:termSetCont}.
\end{rem}

We exploit the preceding result to show recursive feasibility of problem \eqref{OCP} in the following proposition. Because constraint satisfaction is a direct result from recursive feasibility, this result ensures safety of the controller in the uncertain environment.

\begin{prop}\label{prop:RecFeas}
		\emph{(Recursive feasibility)} Let Assumptions \ref{assum:containedPred}, \ref{assum:termRegion} and \ref{assum:termSetCont} hold. If \eqref{OCP} has a solution $\boldsymbol{c_k^{\ast}}$, $\boldsymbol{K_k^{\ast}}$ at time $k$, there is a solution at time $k+1$. Furthermore, input and state constraints are satisfied for the closed-loop system, i.e. applying the input \eqref{equ:inputAtk} to \eqref{equ:DynEq}.
\end{prop}
\begin{pf}
	Based on the solution at time $k$, we choose the candidate solution at time $k+1$
	\begin{equation}
		\begin{aligned}
				&\boldsymbol{\tilde{c}_{k+1}} =
				\\ &\hspace{-2ex}\begin{pmatrix}
				c^*_{1|k}+K^*_{(1,0)|k}o(k)\\
				\vdots\\
				c^*_{N-1|k}+K^*_{(N-1,0)|k}o(k)\\
				\kappa (A^{N}x(k) + \sum\limits_{j=0}^{N-1}(A^{N-j-1}B(c^{\ast}_{j|k}+K^{\ast}_{(j,0)|k}o(k))))
			\end{pmatrix}\hspace{-1ex}
		\end{aligned}
		\label{equ:candidate_c}
\end{equation}
and $\boldsymbol{\tilde{K}_{k+1}}$ with matrices $\tilde{K}_{(i,j)|k+1}$ given by
\begin{equation}
	\begin{aligned}
		\begin{cases}
			K^{\ast}_{(i+1,j+1)|k}& j\leq i\leq N-2\\
			\sum\limits_{n=j}^{N-2}\kappa A^{N-n-2}BK^{\ast}_{(n+1,j+1)|k}& j\leq i=N-1.
		\end{cases}
	\end{aligned}
\label{equ:candidate_K}
\end{equation}
The resulting parameterized input and state sequences from this candidate solution are denoted by $\boldsymbol{\tilde{u}_{k+1}}=\boldsymbol{\tilde{c}_{k+1}}+\boldsymbol{\tilde{K}_{k+1}}\boldsymbol{o_{k+1}}$ and $\boldsymbol{\tilde{x}_{k+1}}=\boldsymbol{A}x(k+1)+\boldsymbol{B}\boldsymbol{\tilde{u}_{k+1}}$ with elements $\tilde{u}_{i|k+1}$ and $\tilde{x}_{i|k}$. We show feasibility of $\boldsymbol{\tilde{c}_{k+1}},\boldsymbol{\tilde{K}_{k+1}}$ in two steps.
	
	(I) First we show input and state constraint satisfaction for $i\in\mathbb{Z}_{[0,N-2]}$ and $i\in\mathbb{Z}_{[0,N-1]}$, respectively.
	For this, we determine the parameterized input from the candidate \eqref{equ:candidate_c} and \eqref{equ:candidate_K} to be
	\begin{equation}
		\begin{aligned}
			&\tilde{u}_{i|k+1} \stackrel{\eqref{equ:inputParam}}{=} \tilde{c}_{i|k+1} + \sum\limits_{j=0}^{i}\tilde{K}_{(i,j)|k+1}o_{j|k+1}\\
			&=  c^{\ast}_{i+1|k} + K^{\ast}_{(i+1,0)|k}o(k)+ \sum\limits_{j=0}^{i}K^{\ast}_{(i+1,j+1)|k}o_{j|k+1}
		\end{aligned}
		\label{equ:Input_prev}
	\end{equation}
	for $i\in \mathbb{Z}_{[0,N-2]}$. At time $k$, the input $u(k)=c^{\ast}_{0|k}+K^{\ast}_{(0,0)|k}o(k)$ is applied, cf. \eqref{equ:inputAtk}.
	As $(\boldsymbol{c_k^{\ast}},\boldsymbol{K_k^{\ast}})\in\Pi_k^{c,K}(x(k))$, applying the optimal parameterized input computed at time $k$ yields input and state constraint satisfaction for $i\in\mathbb{Z}_{[0,N-2]}$ for all 
	$\begin{pmatrix}
		o_{0|k}&\dots &o_{N|k}
	\end{pmatrix}\in\boldsymbol{\mathcal{O}_{k}}$. Using Proposition \ref{prop:ContPred}, input and state constraint satisfaction of the candidate solution for $i\in\mathbb{Z}_{[0,N-2]}$ directly follow for all $\begin{pmatrix}
	o_{0|k+1}&\dots &o_{N|k+1}
\end{pmatrix}\in\boldsymbol{\mathcal{O}_{k+1}}$.
Further, $\tilde{x}_{N-1|k+1}\in\mathcal{X}_f$ for all sequences in $\boldsymbol{\mathcal{O}_{k}}$ follows from analogous arguments. From Assumptions \ref{assum:containedPred} and \ref{assum:termSetCont}, we obtain that $o_{N-1|k+1}\in\mathbb{O}_f$ for all $\begin{pmatrix}
	o_{0|k+1}&\dots &o_{N|k+1}
\end{pmatrix}\in\boldsymbol{\mathcal{O}_{k+1}}$
and, thus,
\begin{equation}
	\tilde{x}_{N-1|k+1}\in\mathcal{X}_f\subseteq\{x\,|\,Fx+Go_{N-1|k+1}\leq g\}.
	\label{equ:state_const_N-1}
\end{equation}
This yields state constraint satisfaction for $i\in\mathbb{Z}_{[0,N-1]}$.

(II) Next, we shown input constraint satisfaction at $i=N-1$ and terminal constraint satisfaction, namely, that $\tilde{u}_{N-1|k+1}\in\mathcal{U}$ and $\tilde{x}_{N|k+1}\in\mathcal{X}_f$. As stated above, $\tilde{x}_{N-1|k+1}\in\mathcal{X}_f$ for all sequences in $\boldsymbol{\mathcal{O}_{k}}$. We can determine the candidate input at $i=N-1$
\begin{equation}
	\begin{aligned}
	\tilde{u}_{N-1|k+1} \stackrel{\eqref{equ:inputAtk}}{=}\tilde{c}_{N-1|k+1}+\underbrace{\sum_{j=0}^{N-1}\tilde{K}_{(N-1,j)|k}o_{j|k+1}}_{=:\tilde{u}^o_{N-1|k}}.
	\end{aligned}
\label{equ:utilde_N-1}
\end{equation}
Using $x(k+1)=Ax(k)+Bc^\ast_{0|k}+BK^\ast_{(0,0)|k}o(k)$ and $\tilde{c}_{i|k+1}= c^*_{i+1|k}+K^*_{(i+1,0)|k}o(k)$ for $i\in\mathbb{Z}_{[0,N-2]}$ due to \eqref{equ:candidate_c}
, we can see that
\begin{equation}
	\begin{aligned}
		\hspace{-1ex}\tilde{c}_{N-1|k+1} = \kappa (A^{N-1}x(k+1)+\sum_{i=0}^{N-2}A^{N-i-2}B\tilde{c}_{i|k+1}).\hspace{-1ex}
		\label{equ:v_tilde}
	\end{aligned}
\end{equation}
Further, we can use the candidate $\tilde{K}_{(N-1,j)}$ from \eqref{equ:candidate_K} to obtain the auxiliary variable defined in \eqref{equ:utilde_N-1}
\begin{equation}
	\begin{aligned}
	\tilde{u}^o_{N-1|k}=\sum_{j=0}^{N-1}\sum_{n=j}^{N-2}\kappa A^{N-n-2}BK^{\ast}_{(n+1,j+1)|k}o_{j|k+1}.
	\end{aligned}
\end{equation}
As the limit of the inner sum is $N-2$, it is zero for $j=N-1$ by our notation. Hence, we can change the limit of the outer sum to $N-2$ as well. Using  $K^{\ast}_{(n+1,j+1)|k}=\tilde{K}_{(n,j)|k+1}$ for $ j\leq n \leq N-2$ from \eqref{equ:candidate_K} and swapping the order of summation, we obtain
\begin{equation}
	\tilde{u}^o_{N-1|k} = \kappa\sum_{n=0}^{N-2}A^{N-n-2}B\sum_{j=0}^{n}\tilde{K}_{(n,j)|k+1}o_{j|k+1}.
	\label{equ:u_o}
\end{equation}
Thus, plugging \eqref{equ:v_tilde} and \eqref{equ:u_o} into \eqref{equ:utilde_N-1} yields
\begin{equation}
	\tilde{u}_{N-1|k+1}=\kappa \tilde{x}_{N-1|k+1}
	\label{equ:TermFeas}
\end{equation}
for any environment trajectory since 
\begin{equation*}
	\begin{aligned}
		&\tilde{x}_{N-1|k+1}=A^{N-1}x(k+1)\\
		&+\sum_{n=0}^{N-1}A^{N-n-2}B\tilde{c}_{n|k+1}+\sum_{j=0}^{n}\tilde{K}_{(n,j)|k+1}o_{j|k+1}.
	\end{aligned}
\end{equation*}
 Because \eqref{equ:state_const_N-1} holds for all environment trajectories in $\boldsymbol{\mathcal{O}_{k}}$, $\tilde{u}_{N-1|k+1}\in\mathcal{U}$ and $\tilde{x}_{N|k+1}=(A+B\kappa)\tilde{x}_{N-1|k+1}\in\mathcal{X}_f$ follow from Assumption \ref{assum:termRegion} together with Proposition \ref{prop:ContPred}.
 
 Recursive feasibility directly implies input and state constraint satisfaction at all times $k$ as $Fx(k)+Go(k)=Fx_{0|k}+Go_{0|k}\leq g$ and $u(k)=c^\ast_{0|k}+K^\ast_{0|k}o(k)=c^\ast_{0|k}+K^\ast_{0|k}o_{0|k}\in\mathcal{U}$ for all times $k$. $\hfill\square$
\end{pf}
We make the following standard assumptions on the cost function \citep{Rawlings17} to show closed-loop convergence to the origin.
\begin{assum}\label{assum:continuity}
	The stage cost $\ell (x,u)$ and the terminal cost $V_f$ are continuous.
\end{assum}\par 
\begin{assum}\label{assum:lowerK}
	There exists a function $\alpha_{\ell}\in \mathcal{K}_{\infty}$ such that $\ell (x,u)\geq \alpha_{\ell}(|x|)$.
\end{assum}\par
\begin{assum}\label{assum:TermCost}
	$V_f$ is positive definite and $V_f(Ax+B\kappa x)-V_f(x)\leq -\ell (x,\kappa x)$ for all $x\in \mathbb{X}_f$.
\end{assum}\par 
\begin{assum}\label{assum:boundedK} There exists a constant $\underline{o}>0$ such that for all times $k\geq 0$ 
	the set $\boldsymbol{\mathcal{O}_{k}}$ contains a ball of radius $\underline{o}$.
\end{assum}\par
We make the technical Assumption \ref{assum:boundedK} that the set $\boldsymbol{\mathcal{O}_{k}}$ has a nonempty interior, the size of which must be uniformly lower bounded. Together with boundedness of $\mathcal{U}$, this results in a (uniform) bound of $\boldsymbol{K^\ast_k}$ for all times $k$.
\begin{lem}\label{lem:boundedK}
	\emph{(Boundedness of $\boldsymbol{K_k}$)} Let Assumption \ref{assum:boundedK} hold, then there exists a constant $\hat{K}>0$ such that $||\boldsymbol{K_k}||<\hat{K}$ for all $(\boldsymbol{c_k},\boldsymbol{K_k})\in\Pi_k^{(c,K)}(x)$. 	
\end{lem}
By the definition of $\boldsymbol{\hat{x}_k}$ and $\boldsymbol{\hat{u}_k}$ in \eqref{OCP:dyn} and \eqref{OCP:input}, the cost function 
strongly depends on the choice of $\boldsymbol{\hat{o}_k}$ which can be understood as a nominal prediction of the environment state. 
Another intuitive choice for $\boldsymbol{\hat{o}_k}$ is the worst case sequence
\begin{equation}
	\boldsymbol{\hat{o}_k}(\boldsymbol{c_k},\boldsymbol{K_k})=\arg \max\limits_{\boldsymbol{\hat{o}_k}\in\boldsymbol{\mathcal{O}_{k}}}J(\boldsymbol{\hat{x}_k},\boldsymbol{\hat{u}_k}).\label{equ:maxCost}
\end{equation}

Using a cost function as in \eqref{equ:maxCost} yields the strongest theoretical results as shown in the following Proposition.

\begin{prop}\label{prop:convergence}
	\emph{(Convergence properties of the closed-loop system)} Let Assumptions \ref{assum:containedPred} - 
	\ref{assum:termSetCont} and \ref{assum:continuity} - 
	\ref{assum:boundedK} hold, and consider any initial state $x(0)$ such that Problem \eqref{OCP} is feasible at time $k=0$.Then, for any choice of $\boldsymbol{\hat{o}_k}\in\boldsymbol{\mathcal{O}_k}$ the closed-loop system converges to a region around the origin. If $\boldsymbol{\hat{o}_k}$ is chosen as in \eqref{equ:maxCost}, the state converges to the origin.
\end{prop}
\begin{pf}
	\emph{Part I (Convergence properties of the nominal cost function):} We show convergence to a region around the origin as a result of the solution to \eqref{OCP}, which we denote by $J^\ast(x,k)=\min_{\boldsymbol{c_k},\boldsymbol{K_k}}J(\boldsymbol{\hat{x}_k},\boldsymbol{\hat{u}_k})$, being
	a Lyapunov function for all $k\geq N$. Due to Proposition \ref{prop:RecFeas} and compactness of $\mathcal{X}$, the state is bounded for $k<N$. Thus, considering only the subsequent steps ($k\geq N$) suffices.
	
	Positiveness of the cost function for all $x\neq 0$ follows directly from Assumption~\ref{assum:lowerK} as
	\begin{equation}
		J^\ast(x,k)\geq \ell(x,c^\ast_{0|k}+K^{\ast}_{(0,0)|k}o(k))\geq \alpha_{\ell}(|x|).
		\label{equ:value_fcn_pos_def}
	\end{equation}
	Let us consider $k\geq N$, then $J^\ast(0,k)=0$ by Assumptions \ref{assum:termRegion}, \ref{assum:termSetCont}, and \ref{assum:TermCost}. Moreover, using compactness of $\mathcal{U}$ and $\mathcal{X}$, continuity of $\ell$ and $V_f$, Assumptions \ref{assum:termRegion}, \ref{assum:termSetCont} and \ref{assum:TermCost}, and following standard arguments \citep[Proposition B.25]{Rawlings17}, we get that
	\begin{equation}
		J^\ast(x,k)\leq \alpha_{up}(|x|)
		\label{equ:value_fcn_K_upper_bound}
	\end{equation}
	for all\footnote{Note that \eqref{equ:value_fcn_K_upper_bound} does not necessarily hold for $k<N$, since the constraints induced by the environment might require the system to initially leave the origin even if $x(0)=0$.} $x\in\mathcal{X}$ and $k\geq N$ for some $\alpha_{up}\in\mathcal{K}$. It remains to show a descent property for $J^\ast$.
	
	Starting from the solution at time $k$, $\boldsymbol{c_k^{\ast}}$ and $\boldsymbol{K_k^{\ast}}$, we can determine the cost of the candidate from Proposition \ref{prop:RecFeas} as in \eqref{equ:candidate_c} and \eqref{equ:candidate_K}, denoted by $\boldsymbol{\tilde{c}_{k+1}}$ and $\boldsymbol{\tilde{K}_{k+1}}$. The $i$-th prediction step of $\boldsymbol{\hat{x}_k}$ and $\boldsymbol{\hat{u}_k}$ is 
	\begin{equation}
		\hat{x}_{i|k} = P^x_i \boldsymbol{\hat{x}_k},\quad \hat{u}_{i|k} = P^u_i \boldsymbol{\hat{u}_k}
	\end{equation}
where $P^x_i$ and $P^u_i$ are appropriate projection matrices.
	The candidate solution has a cost that depends on $\boldsymbol{\hat{o}_{k+1}}$. Using the environment dynamics yields \begin{equation}
		\boldsymbol{\hat{o}_{k+1}}=\boldsymbol{T}(To(k)+Sv(k))+\boldsymbol{S}\boldsymbol{\hat{v}_{k+1}}
	\end{equation}
	with some input sequence $\boldsymbol{\hat{v}_{k+1}}$ and matrices $\boldsymbol{T}$ and $\boldsymbol{S}$ analogous to $\boldsymbol{A}$ and $\boldsymbol{B}$. We can bound the difference between the $i-1$-th state of $\boldsymbol{\hat{o}_{k+1}}$ and the $i$-th state of $\boldsymbol{\hat{o}_{k}}$ based on the difference between the first $i-1$ inputs of $\begin{pmatrix}v(k)^\top&\boldsymbol{\hat{v}_{k+1}}^\top \end{pmatrix}^\top$ and $\boldsymbol{\hat{v}_{k}}$. By the standing Assumption on $\mathcal{V}$, this difference between the inputs is bounded by $\bar{v}$.
	
	We define $\boldsymbol{\tilde{u}_{k+1}}=\boldsymbol{\tilde{c}_{k+1}}+\boldsymbol{\tilde{K}_{k+1}}\boldsymbol{\hat{o}_{k+1}}$ and $\boldsymbol{\tilde{x}_{k+1}}=\boldsymbol{A}(Ax(k)+B(c^\ast_{0|k}+K^\ast_{(0,0)|k}o(k)))+\boldsymbol{B}\boldsymbol{\tilde{u}_{k+1}}$. From the definitions of $\boldsymbol{\tilde{c}_{k+1}}$ and $\boldsymbol{\tilde{K}_{k+1}}$ in \eqref{equ:candidate_c} and \eqref{equ:candidate_K} and Assumption \ref{assum:containedPred}, it can be seen that for all $i\in\mathbb{Z}_{[1,N]}$
	\begin{equation}\label{equ:bound_uk_uk+1}
		||P^u_i\boldsymbol{\hat{u}_k}-P^u_{i-1}\boldsymbol{\tilde{u}_{k+1}}||\leq  i||P^u_i\boldsymbol{K^\ast_k}\boldsymbol{S}||\bar{v}
	\end{equation}
and 
	\begin{equation}\label{equ:bound_xk_xk+1}
	||P^x_i\boldsymbol{\hat{x}_k}-P^x_{i-1}\boldsymbol{\tilde{x}_{k+1}}||\leq i||P^x_i\boldsymbol{BK^\ast_k}\boldsymbol{S}||\bar{v}
.
\end{equation}
The cost of the candidate solution can be determined as
	\begin{equation}
		\begin{aligned}
	&\tilde{J}_{k+1}:=J(\boldsymbol{\tilde{x}_{k+1}},\boldsymbol{\tilde{u}_{k+1}})\\
	&=\sum\limits_{i=0}^{N-1}\ell (P^x_i \boldsymbol{\tilde{x}_{k+1}},P^u_i \boldsymbol{\tilde{u}_{k+1}})+V_f(P^x_N \boldsymbol{\tilde{x}_{k+1}}).\label{equ:candCost}
	\end{aligned}
\end{equation}
Due to Assumption \ref{assum:continuity} and the boundedness of input and state constraints
, there are $\mathcal{K}$-functions $\alpha_{1},\alpha_{2}$ such that
\begin{equation}
	\ell(x+\delta x,u+\delta u)\leq \ell (x,u)+ \alpha_1(||\delta x||+||\delta u||)\label{equ:l_cont}
\end{equation}
 for any $x+\delta x\in\mathcal{X}$ and $u+\delta u\in\mathcal{U}$ and
\begin{equation}
	V_f(x+\delta x)\leq V_f(x)+ \alpha_2(||\delta x||)
	\label{equ:V_f-cont}
\end{equation}
for any $x+\delta x\in\mathcal{X}_f$ \citep{Rawlings15}. With Assumption \ref{assum:TermCost}, \eqref{equ:TermFeas} yields
\begin{equation}\label{equ:term_desc}
	\begin{aligned}
	&	\ell (P^x_{N-1}\boldsymbol{\tilde{x}_{k+1}},P^u_{N-1}\boldsymbol{\tilde{u}_{k+1}})+V_f(P^x_N \boldsymbol{\tilde{x}_{k+1}})\\
	&	\leq V_f(P^x_{N-1} \boldsymbol{\tilde{x}_{k+1}}).
	\end{aligned}
\end{equation} 
Thus, combining \eqref{equ:bound_uk_uk+1}, \eqref{equ:bound_xk_xk+1}, \eqref{equ:candCost}, \eqref{equ:l_cont}, \eqref{equ:V_f-cont}, and \eqref{equ:term_desc}, there is an upper bound for the candidate cost 
	\begin{equation}
		\begin{aligned}
	\tilde{J}_{k+1}\leq&
	\sum\limits_{i=1}^{N-1}\ell (P^x_i \boldsymbol{\hat{x}_{k}},P^x_i \boldsymbol{\hat{u}_{k}})+V_f (P^x_N\boldsymbol{\hat{x}_{k}},P^u_{N}\boldsymbol{\hat{u}_{k}})\\&+\alpha_{Kv} (||\boldsymbol{K^\ast_k}\boldsymbol{S}||\bar{v})
	\end{aligned}\label{equ:candUpperBound}
\end{equation}
with 
\begin{equation*}
		\begin{aligned}
			\hspace{-1ex}\alpha_{Kv} (s):=\hspace{-0.5ex}\alpha_2( N||P^x_i\boldsymbol{B}||s)+
			\sum\limits_{i=1}^{N-1}\alpha_1(i(||P^u_i|| + ||P^x_i\boldsymbol{B}||)s)
			.\hspace{-1ex}
			\end{aligned}
	\end{equation*}
From Lemma \ref{lem:boundedK}, it holds that $||\boldsymbol{K_k}||\leq \hat{K}$ for some constant $\hat{K}\geq 0$. Thus, we can define $\alpha_{v}(\bar{v}):= \alpha_{Kv}(\hat{K}||\boldsymbol{S}||\bar{v})$. Using Assumption \ref{assum:lowerK}, Lemma \ref{lem:boundedK} and the principle of optimality, we can finally state that
\begin{equation}
	J^\ast(x^+,k+1)\leq \tilde{J}_{k+1}\leq J^\ast(x,k)+\alpha_v(\bar{v})-\alpha_{\ell}(x(k)),\label{equ:descent_nominal}
\end{equation}
which combined with \eqref{equ:value_fcn_pos_def} and \eqref{equ:value_fcn_K_upper_bound} yields convergence to a region around the origin using standard Lyapunov arguments \citep[Lemma 3.5]{Jiang01}. 
 
\emph{Part II (Convergence properties with the min-max cost function):} Using the worst case environment prediction as in \eqref{equ:maxCost} is equivalent to solving a min-max problem. Eqns. \eqref{equ:value_fcn_pos_def} and \eqref{equ:value_fcn_K_upper_bound} also hold for the min-max problem by analogous arguments. We can construct a feasible candidate from the solution of the min-max problem at time $k$ analogously to \eqref{equ:candidate_c} and \eqref{equ:candidate_K}. Since the candidate in \eqref{equ:candidate_c} and \eqref{equ:candidate_K} is independent of $\boldsymbol{\hat{o}_{k+1}}$, we can determine the cost of the candidate to be the same as \eqref{equ:candCost} with worst case $\boldsymbol{\hat{o}_{k+1}^\ast}(\boldsymbol{\tilde{c}_{k+1}},\boldsymbol{\tilde{K}_{k+1}})$ as in \eqref{equ:maxCost}. Only performing the inner maximization for the candidate yields a cost $\tilde{J}_{k+1}$ that is greater than or equal to the solution of the min-max problem at time $k+1$, $J^\ast_{k+1}$. Concatenating $o(k)$ and the first $N-1$ steps of $\boldsymbol{\hat{o}_{k+1}^\ast}$, yields a sequence $\boldsymbol{\bar{o}_k}\in\boldsymbol{\mathcal{O}_{k}}$ by Proposition \ref{prop:ContPred}. Computing the cost of this trajectory with the solution of the min-max problem at time $k$, $\boldsymbol{c^\ast_k}$ and $\boldsymbol{K^\ast_{k}}$, results in a cost $\bar{J}_k$ that is smaller than the min-max cost $J^\ast_k$. Further, the following relation holds between $\tilde{J}_{k+1}$ and $\bar{J}_{k}$ by Assumptions \ref{assum:lowerK} and \ref{assum:TermCost},
 \begin{equation}
 	\begin{aligned}
 	&\tilde{J}_{k+1} = \bar{J}_{k} \\
 	&+ V_f(P_N^x(\boldsymbol{A}x(k+1)+\boldsymbol{B}(\boldsymbol{\tilde{c}_{k+1}}+\boldsymbol{\tilde{K}_{k+1}}\boldsymbol{\hat{o}^\ast_{k+1}})))\\
 	&-V_f(P_{N-1}^x(\boldsymbol{A}x(k+1)+\boldsymbol{B}(\boldsymbol{\tilde{c}_{k+1}}+\boldsymbol{\tilde{K}_{k+1}}\boldsymbol{\hat{o}^\ast_{k+1}})))\\
 	&+\ell(P_{N-1}^x(\boldsymbol{A}x(k+1)+\boldsymbol{B}(\boldsymbol{\tilde{c}_{k+1}}+\boldsymbol{\tilde{K}_{k+1}}\boldsymbol{\hat{o}^\ast_{k+1}})),\\
 	&\kappa P_{N-1}^x(\boldsymbol{A}x(k+1)+\boldsymbol{B}(\boldsymbol{\tilde{c}_{k+1}}+\boldsymbol{\tilde{K}_{k+1}}\boldsymbol{\hat{o}^\ast_{k+1}})))\\
 	&-\ell(x(k),c^\ast_{0|k}+K^\ast_{(0,0)|k}o(k)) \leq \bar{J}_{k} -\alpha_{\ell}(|x(k)|).
 	\label{equ:decent_worst_case}
 	\end{aligned} 
\end{equation} 
Thus, we conclude that
\begin{equation}
	J^{\ast}_{k+1}\leq \tilde{J}_{k+1}\leq \bar{J}_{k} -\alpha_{\ell}(|x(k)|)\leq J^\ast_k -\alpha_{\ell}(|x(k)|),
\end{equation}
which implies convergence to the origin by standard Lyapunov arguments and concludes the proof.  
 $\hfill\square$
	\end{pf}
Proposition \ref{prop:convergence} guarantees asymptotic convergence with the worst-case cost \eqref{equ:maxCost}, but only convergence to a region around the origin for an arbitrary environment prediction $\boldsymbol{\hat{o}_k} \in \boldsymbol{\mathcal{O}_{k}}$. Using the worst case cost function would mean solving a min-max problem online which is computationally expensive or intractable in many cases and, thus, prohibitive. In the nominal case, the region to which the closed-loop system state converges depends on $\bar{v}$, which relates to the maximum possible deviation between consecutive environment trajectory predictions. This results from the fact that in case of using the worst-case cost \eqref{equ:maxCost}, the value function decreases at each time step as shown in \eqref{equ:decent_worst_case}, whereas this does not need to be the case for an arbitrary environment prediction $\boldsymbol{\hat{o}_k} \in \boldsymbol{\mathcal{O}_{k}}$ at each time k, compare \eqref{equ:descent_nominal}. A more sophisticated choice of $\boldsymbol{\hat{o}_k}$ or a modified cost function may lead to the same convergence property as choosing the worst-case cost function.
\begin{rem}
	The problem as formulated in \eqref{OCP} contains a semi-infinite constraint due to the definition of $\Pi^{c,K}_k(x)$. We can resolve this semi-infinite constraint by making use of Lagrange duality of the linear constraints analogously to \cite{Goulart06}. \label{rem:impl}
	While this procedure achieves a computationally tractable reformulation of the semi-infinite constraint, this comes at the cost of additional auxiliary decision variables. 
	The number of auxiliary decision variables scales quadratically with the prediction horizon yielding a rather high computational complexity.
	\end{rem}
\section{Simulation example}
We consider a platoon of three vehicles approaching a red light in the same lane. One vehicle is controlled and is surrounded by two vehicles which are driven by humans. The human driven vehicles constitute the environment. The constraints are defined by a minimum distance that has to be kept to the leading and the following vehicle and a minimum time-to-collision (TTC) denoted by $\tau$. The TTC describes how long it would take for two vehicles to collide if both keep their velocity constant. An additional constraint is that the controlled vehicle has to maintain a non-negative velocity. Moreover, we assume that there is a target position that can be reached within the prediction horizon of $N = 12$ steps and that the two surrounding vehicles behave accordingly. Further, the two environment vehicles cannot drive backwards and, for the sake of this scenario, maintain a minimum distance between each other as well as a maximal velocity difference.

\begin{figure}
	\centering
	\begin{subfigure}[t]{0.5\linewidth}
		\centering
		\scalebox{0.33}[0.4]{
			\includegraphics{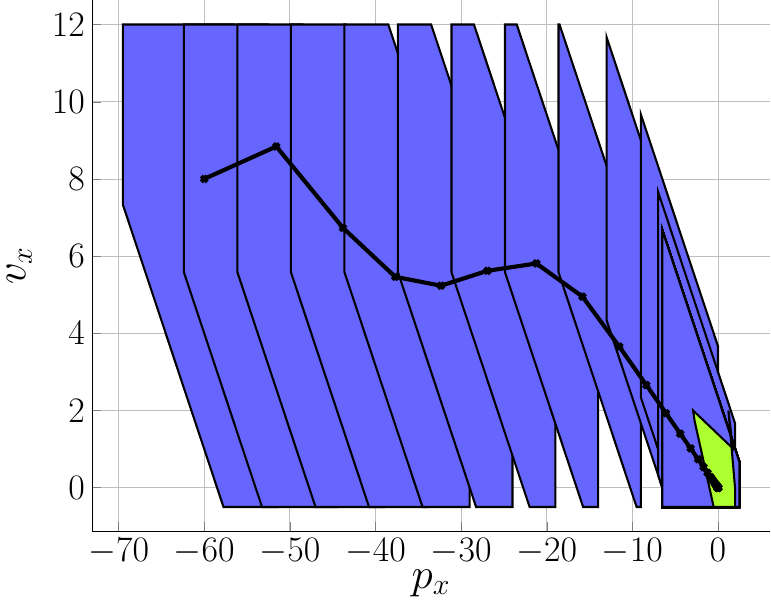}}
		\caption{Closed-loop simulation with environment A.}
		\label{fig:Scen1}
	\end{subfigure}
\hfill
	\begin{subfigure}[t]{0.48\linewidth}
		\centering
		\scalebox{0.33}[0.4]{
			\includegraphics{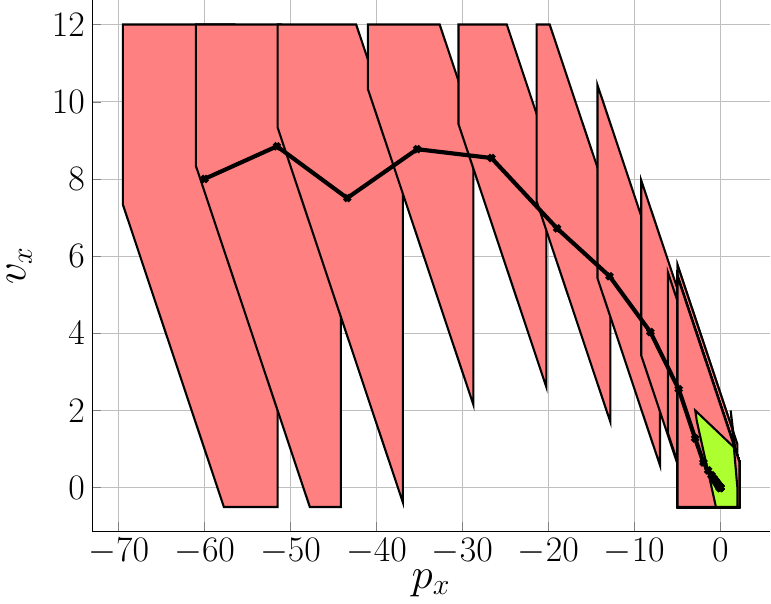}}
		\caption{Closed-loop simulation with environment B.}
		\label{fig:Scen2}
	\end{subfigure}
	\hfill
	\begin{subfigure}[t]{0.48\linewidth}
	\centering
	\scalebox{0.3}[0.3]{
		\includegraphics{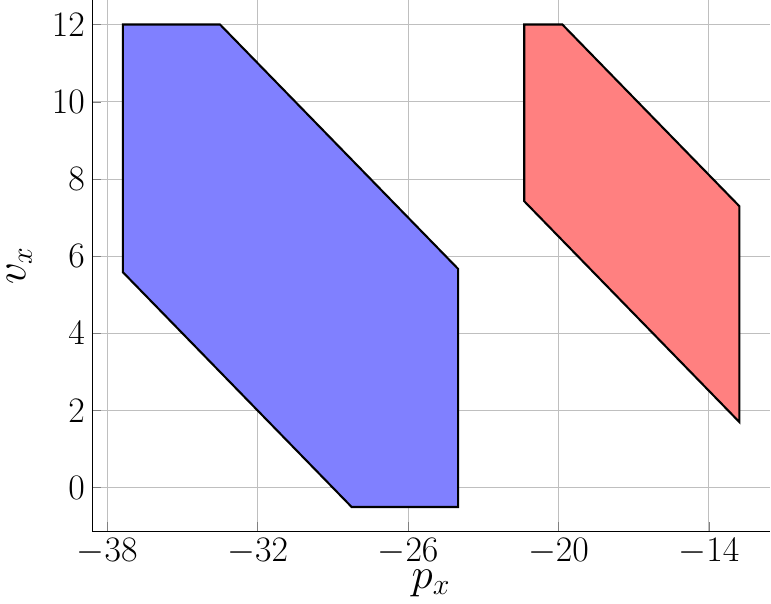}}
	\caption{The constraint sets for $k=5$ are non-intersecting.}
	\label{fig:ConstrSet}
\end{subfigure}
\hfill
\begin{subfigure}[t]{0.48\linewidth}
	\centering
	\scalebox{0.3}[0.3]{
		\includegraphics{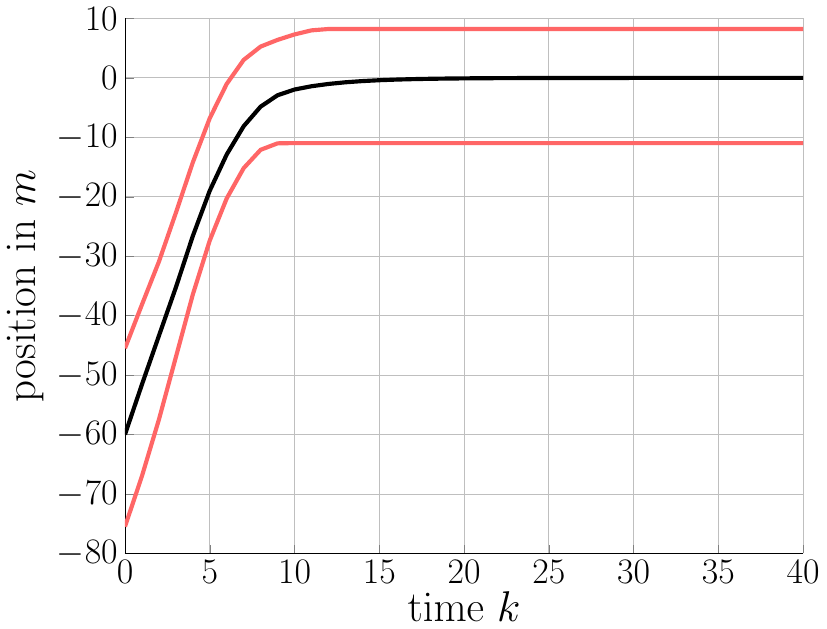}}
	\caption{Closed-loop position of controlled vehicle (black) and human-driven vehicles (red) from environment B.}
	\label{fig:timeseries}
\end{subfigure}
	\caption{\small Closed-loop simulation results for two different environment trajectories (A and B). The state trajectories are given in black and the actual constraint sets in blue and red, respectively. The terminal region is depicted in green.}
	\label{fig:SimRes}
\end{figure}

The controlled vehicle is modeled by a double integrator
\begin{equation}
	x(k+1)=\begin{pmatrix}
		1&\delta\\
		0&1\\
	\end{pmatrix}x(k)+\begin{pmatrix}
		\delta^2/2\\
		\delta
	\end{pmatrix}u(k)
\end{equation}
with position and velocity as states, $x(k)=(
	p_x(k),v_x(k)
)^\top$, and acceleration as the input, $u(k)=a_x(k)$. The two environment vehicles are represented by two double integrators
 \begin{equation}
 	o(k+1)=\begin{pmatrix}
 		1&\delta&0&0\\
 		0&1&0&0\\
 		0&0&1&\delta\\
 		0&0&0&1
 	\end{pmatrix}o(k)+\begin{pmatrix}
 		\delta^2/2&0\\
 		\delta&0\\
 		0&\delta^2/2\\
 		0&\delta
 	\end{pmatrix}v(k)
 \end{equation}
with positions and velocities as states \begin{equation*}
	o(k) =(
	p_{o,1}(k),v_{o,1}(k), p_{o,2}(k),v_{o,2}(k)
	)^\top
\end{equation*} and accelerations as inputs, 
\begin{equation*}
	v(k)=(
	a_{o,1}(k),a_{o,2}(k)
	)^\top.
\end{equation*}
Let the sampling time be $\delta = 1\, s$. The above described state constraints are given by $\mathcal{X}=\{x\,|\, F x+G o \leq g\}$ with
\begin{equation}
\hspace{-1.5ex}	F = \begin{pmatrix}
		1&0\\
		-1&0\\
		0&1\\
		0&-1\\
		1&\tau\\
		-1&-\tau\\
	\end{pmatrix}\hspace{-0.6ex},G  = \begin{pmatrix}
		-1&0&0&0\\
		0&0&1&0\\
		0&0&0&0\\
		0&0&0&0\\
		-1&-\tau&0&0\\
		0&0&1&\tau\\
	\end{pmatrix}\hspace{-0.6ex},g = \begin{pmatrix}
-6\, m\\ -6\, m\\ 12\, ms^{-1}\\0.05\, s^{-1}\\-5\, m\\-5\, m
\end{pmatrix}\hspace{-1.2ex}
\end{equation}
and $\tau = 1.5\, s$. The prediction sets of the environment vehicle $\mathcal{O}_{i|k}$ incorporate that the velocities are positive and that there is a minimum distance of $12\, m$ and TTC of $2.5\,\tau$ between the environment vehicles for the first $N-1$ time steps and that each vehicle stays within a target region for all subsequent steps ($k\geq N$). The acceleration of the vehicles is between $-2$ and $1\, ms^{-1}$. It is easy to verify that the environment prediction satisfies Ass.~\ref{assum:containedPred} because the sets $\mathcal{V}_{i|k}$ are constant and we have $\mathcal{O}_{i|k}=\mathcal{O}_{i-1|k+1}$ for all $i$ and $k$. We choose a quadratic cost function $\ell(x,u)= ||x||^2_Q +||u||^2_R$, the terminal controller as the linear quadratic regulator for the chosen $Q\succ 0$ and $R\succ 0$ and the terminal cost as $V_f(x)=||x||_P$ with $P\succ 0$ being the solution of the algebraic Ricatti equation. The terminal region is depicted in green in Fig.~\ref{fig:Scen1} and \ref{fig:Scen2} and is designed such that it is invariant for the terminal controller (Ass.~\ref{assum:termRegion}) and such that the constraints are satisfied for all environment states in the target region (Ass.\ref{assum:termSetCont}).

We performed the simulations using Octave and the CasADi Toolbox 
 \citep{Andersson2019}. The initial conditions of the vehicles are given by $x(0)=(-60,8)^\top$ and $o(0)=(-45.5,8,-75.5,8)^\top$. To illustrate the constraint evolution, we first sample two possible sequences of the environment vehicles (corresponding to, loosely speaking, fast and slow human driven vehicles) that yield non-intersecting constraint sets depicted at $k=5$ in Fig.~\ref{fig:ConstrSet}. As discussed in the introduction, such a large uncertainty in the environment cannot be handled by existing approaches (compare Fig. 1). The resulting closed-loop trajectories and the constraint sets from the respective environment sequence are given in Fig.~\ref{fig:Scen1} and Fig.~\ref{fig:Scen2}. In Fig.~\ref{fig:timeseries}, we show the closed-loop positions of the three vehicles for one of the scenarios.

Our simulation demonstrates the theoretical results, namely satisfaction of the uncertain state constraints and convergence to a region around the origin.
\section{Conclusion}
We presented an input parameterization for robust MPC to guarantee constraint satisfaction in uncertain environments. Namely, we optimize online over policies that include a feedback term from the environment. This allows to handle more general cases of uncertain constraints than in existing literature. Furthermore, we proved recursive feasibility and constraint satisfaction under suitable assumptions.

{\footnotesize\bibliography{ifacconf}    }      

\begin{thebibliography}{18}
\providecommand{\natexlab}[1]{#1}
\providecommand{\url}[1]{\texttt{#1}}
\providecommand{\urlprefix}{URL }
\expandafter\ifx\csname urlstyle\endcsname\relax
  \providecommand{\doi}[1]{doi:\discretionary{}{}{}#1}\else
  \providecommand{\doi}{doi:\discretionary{}{}{}\begingroup
  \urlstyle{rm}\Url}\fi

\bibitem[{Althoff and Magdici(2016)}]{Althoff16}
Althoff, M. and Magdici, S. (2016).
\newblock Set-based prediction of traffic participants on arbitrary road
  networks.
\newblock \emph{IEEE Transactions on Intelligent Vehicles}, 1(2), 187--202.

\bibitem[{Ames et~al.(2019)Ames, Coogan, Egerstedt, Notomista, Sreenath, and
  Tabuada}]{Ames19}
Ames, A., Coogan, S.D., Egerstedt, M., Notomista, G., Sreenath, K., and
  Tabuada, P. (2019).
\newblock Control barrier functions: Theory and applications.
\newblock \emph{2019 18th European Control Conference (ECC)}, 3420--3431.

\bibitem[{Andersson et~al.(2019)Andersson, Gillis, Horn, Rawlings, and
  Diehl}]{Andersson2019}
Andersson, J.A.E., Gillis, J., Horn, G., Rawlings, J.B., and Diehl, M. (2019).
\newblock {CasADi} -- {A} software framework for nonlinear optimization and
  optimal control.
\newblock \emph{Mathematical Programming Computation}, 11(1), 1--36.
\newblock \doi{10.1007/s12532-018-0139-4}.

\bibitem[{Batkovic et~al.(2023{\natexlab{a}})Batkovic, Ali, Falcone, and
  Zanon}]{Batkovic23}
Batkovic, I., Ali, M., Falcone, P., and Zanon, M. (2023{\natexlab{a}}).
\newblock Safe trajectory tracking in uncertain environments.
\newblock \emph{IEEE Transactions on Automatic Control}, 68(7), 4204--4217.
\newblock \doi{10.1109/TAC.2022.3207875}.

\bibitem[{Batkovic et~al.(2023{\natexlab{b}})Batkovic, Gupta, Zanon, and
  Falcone}]{Batkovic23b}
Batkovic, I., Gupta, A., Zanon, M., and Falcone, P. (2023{\natexlab{b}}).
\newblock Experimental validation of safe mpc for autonomous driving in
  uncertain environments.
\newblock \emph{IEEE Transactions on Control Systems Technology}.

\bibitem[{Goulart et~al.(2006)Goulart, Kerrigan, and Maciejowski}]{Goulart06}
Goulart, P.J., Kerrigan, E.C., and Maciejowski, J.M. (2006).
\newblock Optimization over state feedback policies for robust control with
  constraints.
\newblock \emph{Automatica}, 42(4), 523--533.

\bibitem[{Jiang and Wang(2001)}]{Jiang01}
Jiang, Z.P. and Wang, Y. (2001).
\newblock Input-to-state stability for discrete-time nonlinear systems.
\newblock \emph{Automatica}, 37, 857--869.
\newblock \doi{10.1016/S0005-1098(01)00028-0}.

\bibitem[{Kolmanovsky et~al.(2014)Kolmanovsky, Garone, and
  Di~Cairano}]{Kolmanovsky14}
Kolmanovsky, I., Garone, E., and Di~Cairano, S. (2014).
\newblock Reference and command governors: A tutorial on their theory and
  automotive applications.
\newblock In \emph{2014 American Control Conference}, 226--241.
\newblock \doi{10.1109/ACC.2014.6859176}.

\bibitem[{Langson et~al.(2004)Langson, Chryssochoos, Raković, and
  Mayne}]{Langson04}
Langson, W., Chryssochoos, I., Raković, S., and Mayne, D. (2004).
\newblock Robust model predictive control using tubes.
\newblock \emph{Automatica}, 40(1), 125--133.

\bibitem[{Liu and Stursberg(2019)}]{Liu19}
Liu, Z. and Stursberg, O. (2019).
\newblock Recursive feasibility and stability of mpc with time-varying and
  uncertain state constraints.
\newblock In \emph{2019 18th European Control Conference (ECC)}, 1766--1771.

\bibitem[{Manrique et~al.(2014)Manrique, Fiacchini, Chambrion, and
  Millerioux}]{Manrique14}
Manrique, T., Fiacchini, M., Chambrion, T., and Millerioux, G. (2014).
\newblock Mpc tracking under time-varying polytopic constraints for real-time
  applications.
\newblock In \emph{2014 European Control Conference (ECC)}, 1480--1485.

\bibitem[{Mayne(2014)}]{Mayne14}
Mayne, D.Q. (2014).
\newblock Model predictive control: Recent developments and future promise.
\newblock \emph{Automatica}, 50(12), 2967--2986.

\bibitem[{Mayne et~al.(2009)Mayne, Raković, Findeisen, and
  Allgöwer}]{Mayne09}
Mayne, D., Raković, S., Findeisen, R., and Allgöwer, F. (2009).
\newblock Robust output feedback model predictive control of constrained linear
  systems: Time varying case.
\newblock \emph{Automatica}, 45(9), 2082--2087.

\bibitem[{Nair et~al.(2022)Nair, Tseng, and Borrelli}]{Nair22}
Nair, S.H., Tseng, E.H., and Borrelli, F. (2022).
\newblock Collision avoidance for dynamic obstacles with uncertain predictions
  using model predictive control.
\newblock In \emph{2022 IEEE 61st Conference on Decision and Control (CDC)},
  5267--5272.
\newblock \doi{10.1109/CDC51059.2022.9993319}.

\bibitem[{Rawlings et~al.(2017)Rawlings, Mayne, and Diehl}]{Rawlings17}
Rawlings, J., Mayne, D., and Diehl, M. (2017).
\newblock \emph{Model Predictive Control: Theory, Computation, and Design}.
\newblock Nob Hill Publishing.

\bibitem[{Rawlings and Risbeck(2015)}]{Rawlings15}
Rawlings, J.B. and Risbeck, M.J. (2015).
\newblock On the equivalence between statements with -$\delta$ and k-functions.
\newblock In \emph{TWCCC Tech. Rep. 2015-01, 2015. [Online]. Available:
  https://engineering.ucsb.edu/jbraw/jbrweb-archives/
  tech-reports/twccc-2015-01.pdf}.

\bibitem[{Robla-Gómez et~al.(2017)Robla-Gómez, Becerra, Llata,
  González-Sarabia, Torre-Ferrero, and Pérez-Oria}]{Robla-Gomez17}
Robla-Gómez, S., Becerra, V.M., Llata, J.R., González-Sarabia, E.,
  Torre-Ferrero, C., and Pérez-Oria, J. (2017).
\newblock Working together: A review on safe human-robot collaboration in
  industrial environments.
\newblock \emph{IEEE Access}, 5, 26754--26773.

\bibitem[{Soloperto et~al.(2019)Soloperto, Köhler, Allgöwer, and
  Müller}]{Soloperto19}
Soloperto, R., Köhler, J., Allgöwer, F., and Müller, M. (2019).
\newblock Collision avoidance for uncertain nonlinear systems with moving
  obstacles using robust model predictive control.
\newblock In \emph{2019 18th European Control Conference (ECC)}, 811--817.
\newblock \doi{10.23919/ECC.2019.8796049}.

\end{thebibliography}
{\renewcommand{\clearpage}{}                                     
\appendix
\vspace{-1ex}
\section{Proof of Lemma~\ref{lem:boundedK}}
\vspace{-1ex}
\begin{pf}
	Per Assumption \ref{assum:boundedK}, there is a sequence $\boldsymbol{o^\circ_k}\in \boldsymbol{\mathcal{O}_{k}}$ such that any sequence $\boldsymbol{o_k}$ with $||\boldsymbol{o^\circ_k}-\boldsymbol{o_k}||\leq \underline{o}$ is contained in $\boldsymbol{\mathcal{O}_{k}}$. From the construction of $\Pi_k^{(c,K)}(x)$, it follows that $\boldsymbol{c_k}+\boldsymbol{K_ko_k}\in\mathcal{U}^{N-1}=\mathcal{U}\times\dots\times \mathcal{U}$ for all $\boldsymbol{o_k}\in\boldsymbol{\mathcal{O}_{k}}$. Thus, 
	 \begin{equation}
	 	\boldsymbol{c_k}+\boldsymbol{K_ko^\circ_k} +\boldsymbol{K_k}(\boldsymbol{o_k}-\boldsymbol{o^\circ_k})\in\mathcal{U}^{N-1}.
	 \end{equation}
 	Since $\boldsymbol{c_k}+\boldsymbol{K_ko^\circ_k}\in\mathcal{U}^N$, we can state that $\boldsymbol{K_k}(\boldsymbol{o_k}-\boldsymbol{o^\circ_k})\in\mathcal{U}^N\oplus(-\mathcal{U}^N)$. Compactness of $\mathcal{U}$ implies that there is a constant $M>0$ such that $||\boldsymbol{K_k}(\boldsymbol{o_k}-\boldsymbol{o^\circ_k})||<M$ for any $||\boldsymbol{o_k}-\boldsymbol{o^\circ_k}||\leq \underline{o}$. Hence, it follows that
 	\begin{equation}
 		\begin{aligned}
 	\hspace{-2ex} ||\boldsymbol{K_k}||= \max_{||\boldsymbol{o_k}-\boldsymbol{o^\circ_k}||= \underline{o}}\frac{||\boldsymbol{K_k}(\boldsymbol{o_k}-\boldsymbol{o^\circ_k})||}{||\boldsymbol{o_k}-\boldsymbol{o^\circ_k}||}  \leq \frac{M}{\underline{o}}=:\hat{K}.
 		\end{aligned}
 	\end{equation}  $\hfill\square$
	\end{pf}}
\end{document}